\documentclass[conference]{IEEEtran}
\IEEEoverridecommandlockouts

\usepackage{cite}
\usepackage{amsmath,amssymb,amsfonts}
\usepackage{algorithmic}
\usepackage{graphicx}
\usepackage{textcomp}
\usepackage{xcolor}
\usepackage{ragged2e}
\usepackage{multirow}
\def\BibTeX{{\rm B\kern-.05em{\sc i\kern-.025em b}\kern-.08em
    T\kern-.1667em\lower.7ex\hbox{E}\kern-.125emX}}
\newcommand{\ts}{\textsuperscript}
\begin{document}

\title{TUNABLE WAVELET UNIT BASED CONVOLUTIONAL NEURAL NETWORK IN OPTICAL COHERENCE TOMOGRAPHY ANALYSIS ENHANCEMENT FOR CLASSIFYING TYPE OF EPIRETINAL MEMBRANE SURGERY
\thanks{This article has been supported in part by UCSD Vision Research Center Core Grant P30EY022589, NIH grant R01EY016323, NIH Grant R01EY033847, National Institutes of Health Bridge2AI Project (OT2OD032644), an unrestricted grant from Research to Prevent Blindness, NY, and unrestricted funds from UCSD Jacobs Retina Center.}
}

\author{\IEEEauthorblockN{An Le\ts{1}, Nehal Mehta\ts{2}, William Freeman\ts{2}, Ines Nagel\ts{2}, Melanie Tran\ts{2}, Anna Heinke\ts{2},\\ Akshay Agnihotri\ts{2}, Lingyun Cheng\ts{2}, Dirk-Uwe Bartsch\ts{2}, Hung Nguyen\ts{1}, Truong Nguyen\ts{1}, Cheolhong An\ts{1}}
\IEEEauthorblockA{\ts{1}Electrical and Computer Engineering Department, University of California San Diego, La Jolla, CA 92093, USA \\
\{d0le,hun004,tqn001,chan\}@ucsd.edu\\
\ts{2}Jacobs Retina Center, Shiley Eye Institute, University of California San Diego, La Jolla, CA 92093, USA\\
\{nnmehta,wrfreeman,inagel,mdtran,aheinke,aagnihotri,l1cheng,dbartsch\}@health.ucsd.edu}
}
\maketitle

\begin{abstract}
In this study, we developed deep learning-based method to classify the type of surgery performed for epiretinal membrane (ERM) removal—either internal limiting membrane (ILM) removal or ERM-alone removal. Our model, based on the ResNet18 convolutional neural network (CNN) architecture, utilizes postoperative optical coherence tomography (OCT) center scans as inputs. We evaluated the model using both original scans and scans preprocessed with energy crop and wavelet denoising, achieving 72\% accuracy on preprocessed inputs, outperforming the 66\% accuracy achieved on original scans. To further improve accuracy, we integrated tunable wavelet units with two key adaptations: Orthogonal Lattice-based Wavelet Units (OrthLatt-UwU) and Perfect Reconstruction Relaxation-based Wavelet Units (PR-Relax-UwU). These units allowed the model to automatically adjust filter coefficients during training and were incorporated into downsampling, stride-two convolution, and pooling layers, enhancing its ability to distinguish between ERM-ILM removal and ERM-alone removal, with OrthLatt-UwU boosting accuracy to 76\% and PR-Relax-UwU increasing performance to 78\%. Performance comparisons showed that our AI model outperformed a trained human grader, who achieved only 50\% accuracy in classifying the removal surgery types from postoperative OCT scans. These findings highlight the potential of CNN based models to improve clinical decision-making by providing more accurate and reliable classifications. To the best of our knowledge, this is the first work to employ tunable wavelets for classifying different types of ERM removal surgery.


\end{abstract}
\section{Introduction}
The retina is a sensory tissue that converts light into neural signals for vision. It consists of ten layers, from the internal limiting membrane (ILM) to the retinal pigment epithelium (RPE) \cite{clinical1}. An epiretinal membrane (ERM), a fibrocellular layer of glial cells, fibroblasts, or macrophages, can form on the ILM, distorting retinal layers and impairing vision \cite{clinical2}. ERM surgery, involving ERM removal with or without ILM peeling, improves retinal structure and function \cite{clinical3}. While ILM removal may reduce ERM recurrence \cite{clinical5}, it is linked to micro-scotomas, reduced retinal sensitivity, and delayed visual recovery \cite{clinical6}, making it a debated topic. Optical coherence tomography (OCT) is commonly used for ERM assessment, but distinguishing ILM removal from postoperative scans remains challenging and often requires surgical notes.

This work explores the use of artificial intelligence (AI) to analyze postoperative OCT scans and determine ILM removal, offering a tool for clinical follow-up and insight into structural differences between surgical techniques. To the best of our knowledge, this is the first study to develop a convolutional neural network (CNN) model for determining ERM surgery types using postoperative OCT scans. ResNet18 \cite{Resnet} was used as the baseline model with center OCT scans as input. To enhance accuracy, we introduced two preprocessing steps: energy-crop and wavelet denoising. Building on prior work \cite{OrthLatt_UwU,AnLe}, we incorporated two tunable wavelet units: OrthoLatt-UwU, enforcing orthogonality and perfect reconstruction through an orthogonal lattice structure, and PR-Relax-UwU, relaxing the perfect reconstruction constraint. These units replaced traditional stride-convolution, downsampling, and pooling functions in ResNet18. Unlike max-pooling \cite{MaxPool} and average-pooling \cite{AveragePool}, which discard fine-grained details, these units preserve both low-pass and high-pass features, improving the model's ability to distinguish between ERM-only and ERM-ILM removal surgeries.
\begin{figure*}[!t]
\centerline{\includegraphics[width=2\columnwidth]{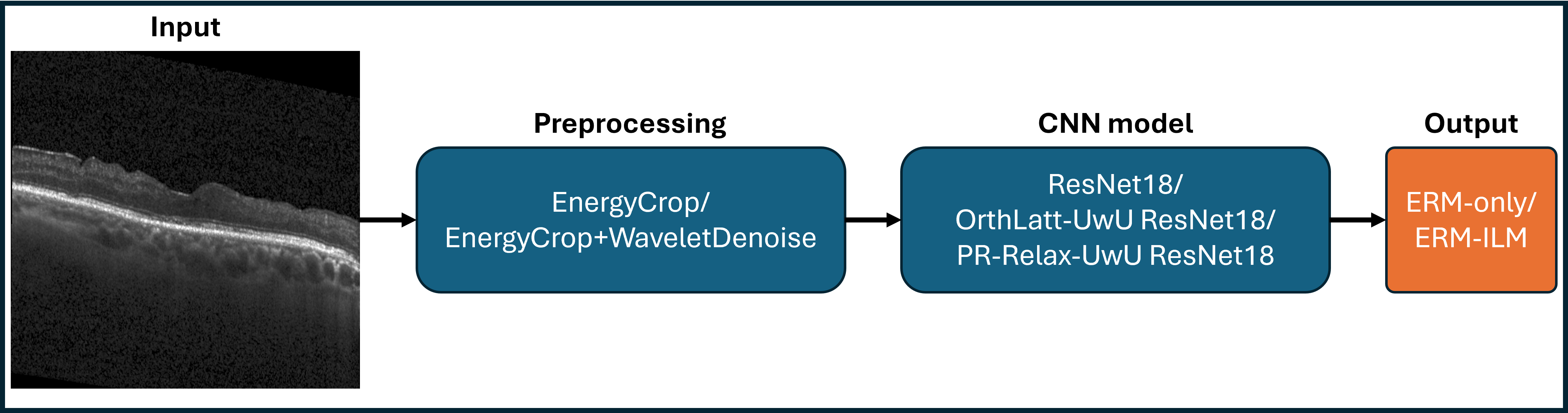}}
\caption{Visualization of the ERM-only/ERM-ILM removal surgery classification pipeline based on OCT center scan input.}
\label{fig:pipeline}
\end{figure*}
AI classifications were compared with those of a trained ophthalmologist, who achieved 50\% accuracy. The best ResNet18 model using original OCT scans reached 66\%, while preprocessing improved accuracy to 72\%. Further integration of OrthoLatt-UwU and PR-Relax-UwU increased accuracy to 76\% and 78\%, respectively. These results demonstrate that CNN-based AI models outperform human graders and highlight the effectiveness of preprocessing and tunable wavelet unit integration in postoperative OCT analysis. This is also the first study to employ tunable wavelets in AI models for medical imaging applications.
\section{Literature Review}
Many studies indicate that ILM removal can alter the inner retinal layers, potentially affecting the integrity and function of remaining retinal cells \cite{clinical13}. Macular surgical techniques continue to advance toward minimizing tissue trauma, ensuring stable anatomical and visual outcomes, and preventing recurrence \cite{clinical14,clinical15}. With AI increasingly integrated into medical practice, its potential to refine surgical techniques is gaining recognition \cite{clinical16}. This capability is particularly useful for detecting subtle OCT changes that human observers might overlook and for patients transitioning to a new physician without access to prior surgical notes. Current research on tissue changes after ERM peeling relies primarily on histological analysis of ex vivo samples, limiting real-time retinal assessment during surgery \cite{clinical22,clinical23}. While histopathology provides valuable post-surgical insights, it is impractical for routine use. In contrast, OCT is widely accessible and commonly used both preoperatively and for follow-up, making it an ideal tool for retinal evaluation \cite{clinical24,clinical25}. Hence, AI-assisted analysis of postoperative OCT scans could help determine whether ERM alone or both ERM and ILM were removed, offering a valuable tool for retina surgeons. However, to the best of our knowledge, no prior studies have employed AI to detect structural differences between ERM-only and ERM-ILM removal using postoperative OCT scans.
\section{Method}
An OCT scan processing pipeline was developed to classify ERM-only and ERM-ILM removal surgeries, as shown in Fig. \ref{fig:pipeline}. The pipeline consists of preprocessing and a CNN-based classification model. Preprocessing includes energy-crop, which removes non-essential regions using vertical pixel energy, and wavelet denoising, which filters high-frequency noise while preserving details. ResNet18 serves as the base CNN model, enhanced with OrthLatt-UwU and PR-Relax-UwU units to retain fine details and high-frequency features, reducing information loss from max-pooling.
\subsection{Preprocessing Procedure}
\begin{figure}[!b]
\centerline{\includegraphics[width=1\columnwidth]{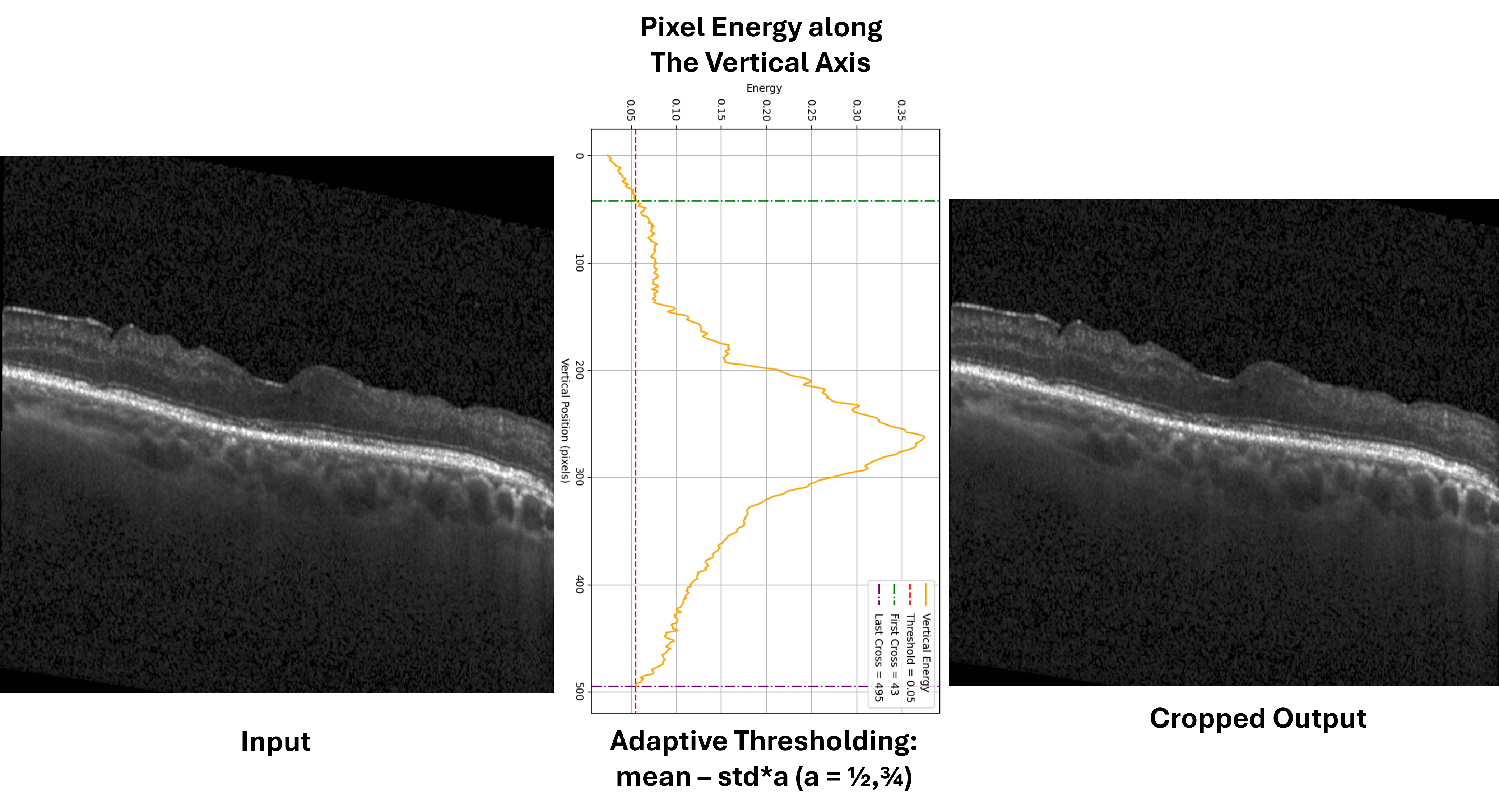}}
\caption{Visualization of the Energy-Crop procedure. The image cropping process is performed using a threshold value defined as $Threshold = mean - a \times StandardDeviation$, where $a$ is the control parameter.}
\label{fig:EnergyCrop}
\end{figure}
\subsubsection{Energy-Crop}
The energy-crop method is designed to remove image areas with little meaningful information, as essential structural details are typically concentrated near the center of an OCT scan. This process uses an adaptive approach to select the relevant area by applying a threshold based on the pixel energy for each vertical coordinate. The pixel energy $E_y$ at a vertical coordinate $y$ is defined as follows:
\begin{equation}
    E_y = \sum_{x}|p_{x,y}|,
\end{equation}
where $p_{x,y}$ is the pixel value at coordinate $(x,y)$ in which $x$ and $y$ represent the horizontal and vertical coordinate indices, respectively. The thresholding value for cropping the image is defined as $Threshold = mean - a \times StandardDeviation$, where $mean$ and $StandardDeviation$ represent the average and standard deviation of all $E_y$, with the controlling parameter $a$. A higher $a$-value results in a smaller cropped area. Multiple $a$-values were tested in order to find the best set-up. Using this threshold, the first and last vertical coordinates where $E_y$ exceeds $Threshold$ are identified to define the cropping boundaries. This process is illustrated in Fig. \ref{fig:EnergyCrop}.
\subsubsection{Wavelet-Denoising}
\begin{figure}[!t]
\centerline{\includegraphics[width=1\columnwidth]{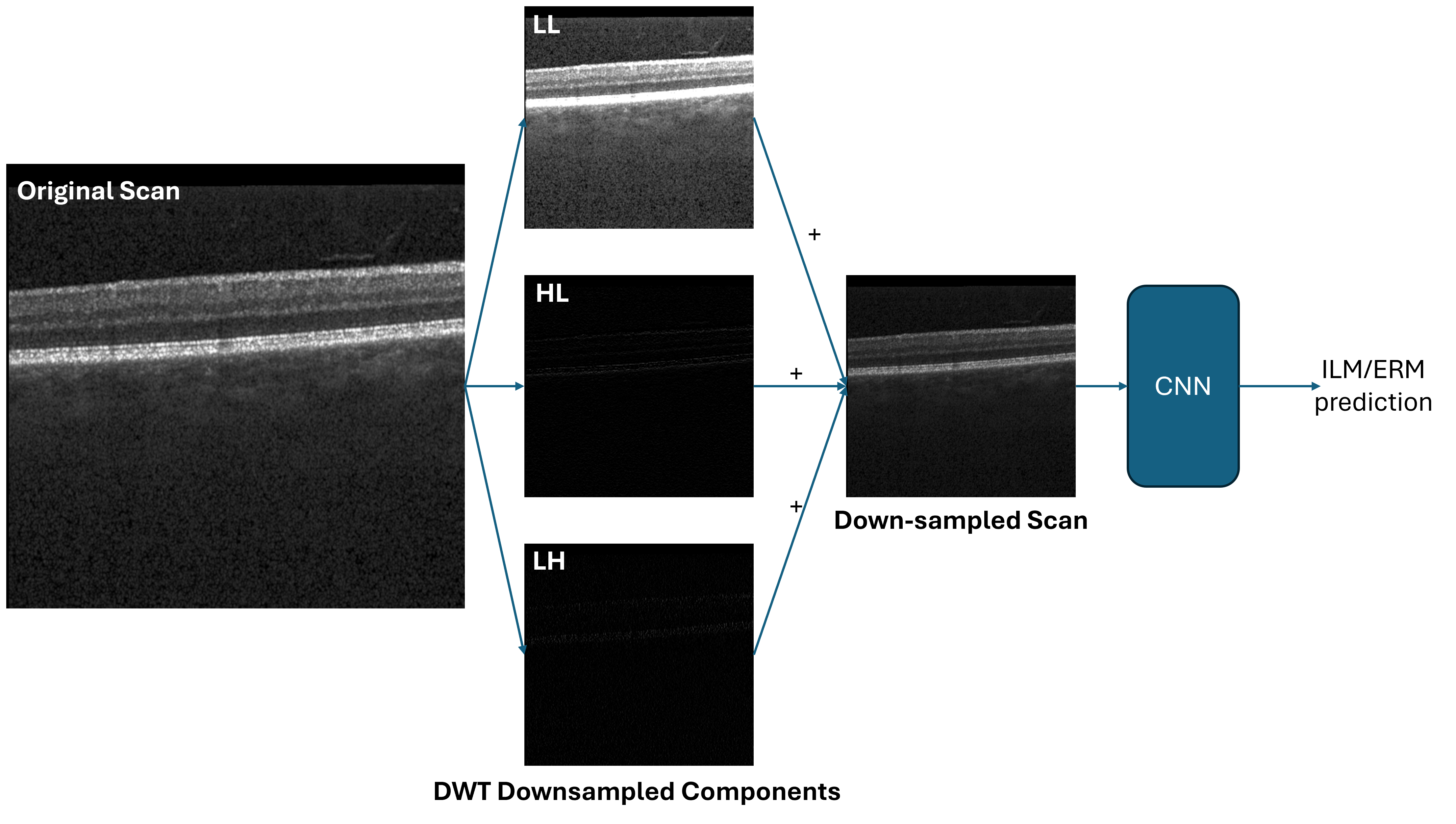}}
\caption{Visualization of the Wavelet Denoising process. The input image is decomposed using Discrete Wavelet Transform (DWT) into an approximation component $LL$, a vertical detail component $LH$, a horizontal detail component $HL$, and a diagonal detail component $HH$. The denoised output is obtained by summing $LL$, $HL$, and $LH$.}
\label{fig:WaveletDenoising}
\end{figure}

The wavelet denoising process is proposed to reduce high-frequency noise while preserving detailed scan features. The process include discrete wavelet transform (DWT) decomposition of the input image into a low-pass component $LL$ and three high-pass components $LH$, $HL$, and $HH$. The denoised image $I_{denoised}$ is defined as the sum of $LL$, $LH$, and $HL$:
\begin{equation}
    I_{denoised} = LL + LH+ HL.
\end{equation}
The process emphasizes the $HL$ and $LH$ components, as they encapsulate the majority of fine structural details present in the OCT scan. In addition, the $HH$ component is discarded, as it contains high-frequency information in both vertical and horizontal orientations, which may represent noise. The process is illustrated in Fig. \ref{fig:WaveletDenoising}.
\subsection{Tunable Wavelet Unit based Models for Surgery Classification}
Tunable wavelet units are integrated into CNN models to enhance classification performance, addressing the loss of fine-grained features caused by max-pooling and average pooling in conventional CNNs. These units employ DWT decomposition with tunable coefficients, enabling the network to retain high-frequency features while optimizing filter coefficients for the classification task.
\subsubsection{Orthogonal Lattice Structure Tunable Wavelet Unit}
In OrthLatt-UwU implementation \cite{OrthLatt_UwU}, the tunable wavelet unit with the low-pass filter $\textbf{H}_{0}$ and high-pass filter $\textbf{H}_{1}$ is constructed with orthogonal lattice structure. The structure can be expressed as follows:
\begin{align}
\label{eq:latticestructure}
 \begin{bmatrix}
   \textbf{H}_{0}(z) \\
   \textbf{H}_{1}(z) \\
   \end{bmatrix} &=  \begin{bmatrix}
   \textbf{H}_{0}(z) \\
   -z^{-L}\textbf{H}_{0}(-z^{-1}) \\
   \end{bmatrix} \\&= \begin{bmatrix} 1&0 \\0&-1 \\ \end{bmatrix}\textbf{R}_{K}\Lambda(z^2)\cdots\textbf{R}_{1}\Lambda(z^2)\textbf{R}_{0}
   \begin{bmatrix}
   1 \\
   z^{-1} \\
 \end{bmatrix},
\end{align}
where $\textbf{R}_{k}$ is a rotation matrix used to construct the filter, with $k = 0, \dots, K$. The delay matrices within the filter are denoted as $\Lambda(z^2)$. Additionally, $N$ represents the order of the filter, defined as $N = 2K + 1$. In addition, rotation matrices are inherently orthogonal. Thus, $\textbf{R}_{k}$ and $\Lambda(z^2)$ can be mathematically expressed as follows:

\begin{gather}
\label{eq:RotationDelayMatrix}
 \textbf{R}_{k} = \begin{bmatrix} cos(\theta_{k})&sin(\theta_{k}) \\-sin(\theta_{k})&cos(\theta_{k}) \\ \end{bmatrix} = \begin{bmatrix} c_k&s_k \\-s_k&c_k \\ \end{bmatrix};\nonumber\\
 \Lambda(z) = \begin{bmatrix} 1&0 \\0&z^{-1} \\ \end{bmatrix}.
\end{gather}
In Eq. \eqref{eq:RotationDelayMatrix}, $\theta_{k}$ represents a rotation angle that determines the coefficients of the wavelet filter bank, with $k = 0, \dots, K$. These rotation angles, used in the rotation matrices—where either their rows or columns are orthonormal—are also referred to as lattice coefficients, defining the filter coefficients within the filter bank. This orthonormality ensures that the resulting filters, obtained through the multiplication of rotation and delay matrices, preserve orthogonality and hence, perfect reconstruction. These lattice coefficients $\theta_{k}$ are also the tunable parameters, making the wavelet unit tunable in CNN.  
\subsubsection{Tunable Wavelet Unit with Perfect Reconstruction Relaxation}
\begin{table*}[!t]
\caption{\justifying{Surgery classification performance based on accuracy with different preprocessing procedures: Energy-Crop (EC) (with $a = 0.75$ and $a = 0.5$) and Wavelet-Denoising (Denoise) (using Haar and DB2 wavelet functions). The baseline performance refers to ResNet18 without preprocessing. The performance of the trained human grader is 50$\%$.}}
\centering
\scalebox{1.35}{
\begin{tabular}{|c|c|c|}
\hline
\multicolumn{1}{|c|}{\textbf{Preprocessing steps}} & \multicolumn{1}{c|}{\textbf{Validation}} & \multicolumn{1}{c|}{\textbf{Test}} \\ \hline
\multicolumn{1}{|c|}{\textbf{Baseline}} & \multicolumn{1}{c|}{$60.29 (\pm2.08 )\%$} & \multicolumn{1}{c|}{$58.8 (\pm7.86 )\%$} \\ \hline
\multicolumn{1}{|c|}{\textbf{EC-0.75}} & \multicolumn{1}{c|}{$64.41 (\pm3.77 )\%$} & \multicolumn{1}{c|}{$62  (\pm6.81)\%$} \\ \hline
\multicolumn{1}{|c|}{\textbf{EC-0.75 + Denoise-Haar}} & \multicolumn{1}{c|}{$67.16 (\pm1.39 )\%$} & \multicolumn{1}{c|}{$61.33 (\pm4.11 )\%$} \\ \hline
\multicolumn{1}{|c|}{\textbf{EC-0.75 + Denoise-DB2}} & \multicolumn{1}{c|}{$64.22 (\pm2.5 )\%$} & \multicolumn{1}{c|}{$64 (\pm2.83 )\%$} \\ \hline

\multicolumn{1}{|c|}{\textbf{EC-0.5}} & \multicolumn{1}{c|}{$61.27 (\pm3.02 )\%$} & \multicolumn{1}{c|}{$62.67 (\pm4.11 )\%$} \\ \hline
\multicolumn{1}{|c|}{\textbf{EC-0.5 + Denoise-Haar}} & \multicolumn{1}{c|}{$61.76 (\pm1.20 )\%$} & \multicolumn{1}{c|}{$62.67 (\pm1.89 )\%$} \\ \hline
\multicolumn{1}{|c|}{\textbf{EC-0.5 + Denoise-DB2}} & \multicolumn{1}{c|}{$62.65 (\pm2.00 )\%$} & \multicolumn{1}{c|}{\textbf{64.67} $(\pm5.25 )\%$} \\ \hline
\end{tabular}}
\label{table:Preprocessing}
\end{table*}
\begin{table*}[!t]
\caption{\justifying{Best surgery classification performances based on accuracy using different proposed methods: Energy-Crop (EC) and Wavelet-Denoising (Denoise) preprocessing procedures, OrthLatt-UwU, and PR-Relax-UwU. The baseline performance refers to ResNet18 without preprocessing. The performance of the trained human grader is 50\%.
}}
\centering
\scalebox{1.35}{
\begin{tabular}{|c|c|c|c|}
\hline
\multicolumn{1}{|c|}{\textbf{Methods}} & \multicolumn{1}{c|}{\textbf{Validation}} & \multicolumn{1}{c|}{\textbf{Test}} & \multicolumn{1}{c|}{\textbf{Best on Test}}\\ \hline
\multicolumn{1}{|c|}{\textbf{Baseline}} & \multicolumn{1}{c|}{$60.29 (\pm2.08 )\%$} & \multicolumn{1}{c|}{$58.8 (\pm7.86 )\%$} & \multicolumn{1}{c|}{$66\%$} \\ \hline
\multicolumn{1}{|c|}{\textbf{EC-0.5}} & \multicolumn{1}{c|}{$61.27 (\pm3.02 )\%$} & \multicolumn{1}{c|}{$62.67  (\pm4.11)\%$} & \multicolumn{1}{c|}{$68\%$} \\ \hline
\multicolumn{1}{|c|}{\textbf{EC-0.5\&Denoise-DB2}} & \multicolumn{1}{c|}{$62.65 (\pm2.00 )\%$} & \multicolumn{1}{c|}{$64.67 (\pm5.25 )\%$} & \multicolumn{1}{c|}{$72\%$} \\ \hline
\multicolumn{1}{|c|}{\textbf{OrthLatt-UwU-8Tap EC-0.5\&Denoise-DB2}} & \multicolumn{1}{c|}{$62.26 (\pm4.86 )\%$} & \multicolumn{1}{c|}{\textbf{69.33 }$(\pm4.99 )\%$} & \multicolumn{1}{c|}{\textbf{76}$\%$}\\ \hline
\multicolumn{1}{|c|}{\textbf{PR-Relax-UwU-8Tap EC-0.5\&Denoise-DB2}} & \multicolumn{1}{c|}{$64.71 (\pm2.40 )\%$} & \multicolumn{1}{c|}{\textbf{70.67} $(\pm5.25 )\%$} & \multicolumn{1}{c|}{\textbf{78}$\%$}\\ \hline
\end{tabular}}
\label{table:Methods}
\end{table*}
The tunable wavelet in the PR-Relax-UwU implementation \cite{AnLe} has $\textbf{H}_{0}$ and $\textbf{H}_{1}$ as the low-pass and high-pass filters, respectively. Since the perfect reconstruction (PR) constraint is relaxed, the filter coefficients satisfy the alias cancellation condition along with a relaxed Half-band condition. Under the satisfied alias cancellation condition, the relationship between $\textbf{H}_{0}$ and $\textbf{H}_{1}$ can be expressed as follows \cite{Wavelets_and_filter_banks}:
\begin{equation}
\label{eq:AliasingCancellation}
\textbf{h}_{1}(n)=(-1)^{n}\textbf{h}_{0}(N-1-n),
\end{equation}
where $\textbf{h}_{0}$, and $\textbf{h}_{1}$ are filter coefficients of $\textbf{H}_{0}$, and $\textbf{H}_{1}$, respectively. In addition, to fulfill the PR constraint, the filter coefficients must satisfy the Half-band condition, which is given as follows \cite{Wavelets_and_filter_banks}:
\begin{equation}
\label{eq:HalfBand}
\textbf{P}(z) + \textbf{P}(-z) = 2,
\end{equation}
where $\textbf{P}(z) = \textbf{H}_{0}(z)\textbf{H}_{0}(z^{-1})$. From (\ref{eq:HalfBand}), the condition for the filter coefficients can be expressed as follows:
\begin{equation}
\label{eq:HalfBand_coef}
\begin{cases} 
\sum_{n=0}^{N-1} {h(n)}^{2} = 1 \\ 
\sum_{0<n,n+2l<N-1}^{N-1} h(n)h(n+2l)=0,\mbox{for }0<l\leq N/2.
\end{cases}
\end{equation}
From (\ref{eq:HalfBand_coef}), the loss function for the PR constraint based on the Half-band condition can be mathematically expressed as follows:
\begin{align}
\label{eq:PRloss}
\textbf{L}_\textbf{PR} =& |1 - \sum_{n=0}^{N-1} h(n)^{2}|^2 + \nonumber\\ &\sum_{l=1}^{N/2} {(\sum_{0<n,n+2l<N-1}^{N-1} h(n)h(n+2l))^2}.
\end{align}
From (\ref{eq:AliasingCancellation}) and (\ref{eq:PRloss}), the PR constraint is implemented to train the filter bank analysis. The relaxation of the PR constraint is achieved by multiplying $\textbf{L}_\textbf{PR}$ with a factor $\alpha$. Cross-Entropy loss $\textbf{L}_\textbf{CE}$ is used. A higher $\alpha$ strengthens the Half-band constraint, while relaxing it allows for greater coefficient fine-tuning. The total loss function $\textbf{L}$ is expressed as follows:
\begin{equation}
\label{eq:Totalloss}
\textbf{L} = \textbf{L}_\textbf{CE} + \alpha\textbf{L}_\textbf{PR}.
\end{equation}

For both tunable wavelet unit cases, given the low-pass and high-pass filters $H_{0}(z)$ and $H_{1}(z)$, the corresponding high-pass and low-pass filter matrices are computed and denoted as $\textbf{H}$ and $\textbf{L}$. These matrices are used to extract the approximation component $\textbf{X}_{ll}$ and the detail components $\textbf{X}_{lh}$, $\textbf{X}_{hl}$, and $\textbf{X}_{hh}$ from an image or feature map $\textbf{X}$. The computation of $\textbf{L}$ can be mathematically described as follows:
\begin{equation}
\label{eq:L_1}
 \textbf{L} = \textbf{D}\widehat{\textbf{H}},
\end{equation}
where $\textbf{D}$ is the downsampling matrix with a factor of 2, and $\widehat{\textbf{H}}$ is a Toeplitz matrix with filter coefficients $h_0, h_1, \dots, h_{2N-1}$ of ${H}_{0}(z)$. The matrix $\textbf{H}$ has a similar structure to $\textbf{L}$ but uses the filter coefficients of ${H}_{0}(z^{-1})$. Using $\textbf{H}$ and $\textbf{L}$, the components $\textbf{X}_{ll}$, $\textbf{X}_{lh}$, $\textbf{X}_{hl}$, and $\textbf{X}_{hh}$ are computed as follows:

\begin{equation}
\label{eq:decomposition}
\begin{aligned}
 \textbf{X}_{ll} &=  \textbf{L}\textbf{X}\textbf{L}^T, &
 \textbf{X}_{lh} &=  \textbf{H}\textbf{X}\textbf{L}^T,\\
 \textbf{X}_{hl} &=  \textbf{L}\textbf{X}\textbf{H}^T, &
 \textbf{X}_{hh} &=  \textbf{H}\textbf{X}\textbf{H}^T.
\end{aligned}
\end{equation}

In this work, we use a one-layer Fully Convolutional Network (FCN) to combine features from the sub-sample low-pass and high-pass components extracted via the Discrete Wavelet Transform (DWT). The combined features can be mathematically expressed as follows:
\begin{align}
\label{eq:OneLayerFCN_formula}
X_{p} = F'(ReLU(\textbf{X}_{ll}),& ReLU(\textbf{X}_{lh}), \nonumber\\ & ReLU(\textbf{X}_{hl}), ReLU(\textbf{X}_{hh})),
\end{align}
where $F'$ is the one-layer FCN with tunable weights. The tunable wavelet units are integrated into the CNN architectures. For the downsampling and pooling layers, we use the decomposed components as inputs for a one-layer Fully Convolutional Network (FCN). In addition, we substitute the two-stride convolution with a non-stride convolution block followed by the tunable wavelet units.
\section{Experiments and Results}

This study adhered to the Declaration of Helsinki and HIPAA regulations. Written informed consent was obtained for surgery, and all patient data were anonymized. OCT scans were reviewed within 12 weeks post-surgery for quality and completeness, with surgical notes determining ILM peeling status. Anonymized SD-OCT scans were imported as video files for human grading and RAW data for AI analysis. The dataset comprises of 340 training-validation samples and 50 test samples, maintaining a balanced 50/50 ERM-only and ERM-ILM distribution. A human grader was trained and tested on a shuffled set (25 eyes per group) without access to surgical notes. The training-validation set was randomly shuffled three times (3-fold experiments) and split 80/20 while preserving the 50/50 balance for the surgery classes. Models were trained on different training-validation splits and evaluated on the same test set, with the trained ophthalmologist achieving 50\% accuracy. All models were trained using the Adam optimizer with a 0.001 learning rate for 200 epochs and a batch size of 128. The preprocessing methods, energy-crop (EC) and wavelet-denoising (Denoise), were evaluated against the baseline ResNet18 model trained on original data. EC experiments tested $a$-values of 0.5 and 0.75, with the best setup applied to Denoise experiments using Haar and DB2 wavelets. Results are reported in Table \ref{table:Preprocessing}.

Based on test set performance in Table \ref{table:Preprocessing}, the best setup (EC-0.5 with Denoise-DB2) was selected for tunable wavelet unit integration experiments with OrthLatt-UwU and PR-Relax-UwU. The tunable wavelet units having filters with 2-Tap (Haar), 4-Tap (DB2), 6-Tap (DB3), and 8-Tap (DB4) were tested, with only the best-performing 8-Tap models reported in Table \ref{table:Methods}. Table \ref{table:Methods} presents the mean accuracies and best test set performance for each method. From Table \ref{table:Preprocessing} and Table \ref{table:Methods}, AI models consistently outperformed trained human graders in distinguishing between ERM-alone and ERM-ILM removal surgeries. The combination of preprocessing and wavelet-based enhancements significantly improved model performance, demonstrating AI's potential in refining postoperative assessment and supporting clinical decision-making. Additionally, relaxing the perfect reconstruction constraint in tunable wavelet unit integration yielded the best classification results.

\section{Conclusion}
This study is the first to develop convolutional neural network (CNN) models for classifying ERM surgery types using postoperative OCT scans. Using ResNet18 as a baseline, we introduced preprocessing techniques, including energy-crop and wavelet downsampling, to improve accuracy. Additionally, tunable wavelet units—OrthoLatt-UwU and PR-Relax-UwU—were integrated to enhance feature extraction and classification. Our results show that AI models consistently outperform trained human graders in distinguishing between ERM-alone and ERM-ILM removal surgeries. The combination of preprocessing and wavelet-based enhancements significantly improved model performance, highlighting AI’s potential in refining postoperative assessment and clinical decision-making. Enhancing OCT analysis with AI offers a non-invasive, cost-effective approach for assessing retinal structure, optimizing surgical techniques, understanding disease pathology, and classifying recurrence. In future work, the current models can be extended to process OCT volumes as inputs instead of single OCT scans, potentially enhancing performance. Additionally, augmentation and diffusion techniques can be employed to increase training data, allowing deeper networks to be trained without overfitting, leading to more stable models with reduced performance variability.

\bibliographystyle{ieeetr}
\bibliography{ref}

\end{document}